\documentclass{IEEEtran}
\usepackage{booktabs}
\usepackage[T1]{fontenc} 
\usepackage{amsmath}
\usepackage{epsfig,endnotes}
\usepackage[colorinlistoftodos,prependcaption,textsize=tiny]{todonotes}
\usepackage{hyperref}

\usepackage{mathtools}
\usepackage{url}
\usepackage[noend]{algpseudocode}
\usepackage{tabularx,colortbl}
\usepackage{multirow}
\usepackage[Symbolsmallscale]{upgreek}
\usepackage{algorithm}
\usepackage{algpseudocode}
\usepackage{mathtools}
\usepackage{url}
\usepackage{authblk}

\title{Game of Drones - Detecting Streamed POI from Encrypted FPV Channel}

\author{
Video - \url{https://www.youtube.com/watch?v=4icQwducz68}
\\ Ben Nassi$^{1}$, Raz Ben-Netanel$^{2}$, Adi Shamir$^{3}$, Yuval Elovici$^{4}$
\\ $^{1}$nassib@post.bgu.ac.il, Dept. of Software and Information Systems Eng., BGU, Beer-Sheva, Israel
\\ $^{2}$razx@post.bgu.ac.il, Dept. of Communication Systems Eng., BGU, Beer-Sheva, Israel
\\ $^{3}$adi.shamir@weizmann.ac.il, Computer Science Department, Weizmann Institute of Science, Rehovot, Israel
\\$^{4}$elovici@bgu.ac.il, Dept. of Software and Information Systems Eng., BGU, Beer-Sheva, Israel
}

\begin{document}
\maketitle
\thispagestyle{empty}

\section*{Abstract}
Drones have created a new threat to people's privacy. We are now in an era in which anyone with a drone equipped with a video camera can use it to invade a subject's privacy by streaming the subject in his/her private space over an encrypted first person view (FPV) channel. Although many methods have been suggested to detect nearby drones, they all suffer from the same shortcoming: they cannot identify exactly what is being captured, and therefore they fail to distinguish between the legitimate use of a drone (for example, to use a drone to film a selfie from the air ) and illegitimate use that invades someone's privacy (when the same operator uses the drone to stream the view into the window of his neighbor's apartment), a distinction that in some cases depends on the orientation of the drone's video camera rather than on the drone's location. In this paper we shatter the commonly held belief that the use of encryption to secure an FPV channel prevents an interceptor from extracting the POI that is being streamed. We show methods that leverage physical stimuli to detect whether the drone's camera is directed towards a target in real time. We investigate the influence of changing pixels on the FPV channel (in a lab setup). Based on our observations we demonstrate how an interceptor can perform a side-channel attack to detect whether a target is being streamed by analyzing the encrypted FPV channel that is transmitted from a real drone (DJI Mavic) in two use cases: when the target is a private house and when the target is a subject.


\section{Introduction}
The proliferation of consumer drones over the last few years \cite{commercial-uav-market-analysis} has created a new privacy threat. We are living in an era in which anyone with a drone equipped with a video camera can invade another individual's privacy by maneuvering the drone to the individual's house and directing the drone's camera to the window of the house in order to film or record the subject in his/her private space. Many privacy invasion incidents have been reported in the media, and laws are being updated to deal with this new threat \cite{Woman-grabs-gun-shoots-nosy-neighbour-s-drone,Virginia-Woman-Shoots-Down-Drone-Near-Actor-Robert-Duvalls-Home,not-my-backyard-man-arrested-after-shooting-drone-down,cheating,case-dismissed-against-william-h-merideth-kentucky-man-arrested-shooting,an-update-on-drone-privacy-concerns,you-cant-shoot-a-drone-so-what-can-you-do-if-it-invades-your-privacy}. 

State of the art drones provide video piloting capabilities, a.k.a. first person view (FPV), a communication channel designed to (1) stream the video captured by the drone's video camera to the operator's controller in order to present the video stream to the operator in real-time, and (2) maneuver the drone by transmitting commands from the controller to the drone. FPV provides excellent infrastructure for a malicious operator to invade someone's privacy without being detected because: (1) it eliminates the need for a malicious operator to be close to the drone or target by allowing the operator to maneuver the drone from far away to a target that is also far away from his/her location, (2) its traffic can be encrypted, and (3) it supports HD resolutions that enable the attacker to obtain high resolution images and close-ups (by using the video camera's zooming capabilities) that are captured by the drone far from the target POI.

Extracting a target POI from an FPV channel has interested researchers for a long time. Many studies have suggested methods for detecting whether a drone is nearby, however none of them can detect exactly what is being captured, and therefore they fail to distinguish between the legitimate use of the drone (e.g., to film a selfie from the air) and illegitimate use (e.g., to stream the view into the window of someone else's apartment), a distinction that in some cases depends on the orientation of the drone's video camera rather than on the drone's location. There are many known cases in which a POI was extracted from an unencrypted video stream \cite{Insurgents-Intercept-Drone-Video,British-and-US-intelligence-hacked-into-Israeli-drones,russia-intercepted-us-drone-over-crimea,Why-Iran's-capture-of-US-drone-will-shake-CIA}, however detecting a target POI from an encrypted video stream remains a challenge.

In this research we shatter the commonly held belief that the use of encryption to secure a surveillance video channel transmitting in real-time prevents an interceptor from extracting the POI that is being tracked by a drone. We present different methods that can be used by an interceptor to detect whether a particular POI (e.g., his/her house, a subject) is being tracked by a drone, even if the FPV channel is encrypted; in order to accomplish this, our proposed methods trigger a physical stimulus that influences the encrypted FPV channel, and can be used by an interceptor to detect if a specific POI is being tracked. We investigate the influence of changing pixels on the FPV channel in a lab setup. Based on our observations we demonstrate how an interceptor can perform a side-channel attack to detect whether a target is being streamed by analyzing the encrypted FPV channel that is transmitted from a real drone (DJI Mavic) in two use cases: when the target is a private house and when the target is a subject. We investigate the amount of time that the physical stimulus must be activated in order to achieve an FPR of zero in each of the use cases.

\subsection{Motivation} 
Interest in intercepting a drone's traffic in order to extract the captured POI from an FPV channel is not limited to civilians; there are many known cases in which a military organization has successfully intercepted an unencrypted surveillance camera's video stream transmitted from a rival's drone and managed to identify the POI that was under surveillance \cite{Insurgents-Intercept-Drone-Video,British-and-US-intelligence-hacked-into-Israeli-drones,russia-intercepted-us-drone-over-crimea,Why-Iran's-capture-of-US-drone-will-shake-CIA}. Just recently, the Israel Defense Forces confirmed that 12 soldiers were killed in 1997 as a result of intercepted intelligence that was transmitted from an Israeli surveillance drone over an unencrypted channel, giving Hezbollah advance knowledge of a naval commando operation deep inside Lebanon \cite{Nasrallah-Ynet,Nasrallah-Ynet2}. However, to the best of our knowledge, there are no known cases in which interceptors were able to extract a POI from an encrypted FPV channel.

\subsection{Contributions}
This study makes the following contributions:
\begin{itemize}

\item \textbf{Extraction of a POI from an encrypted FPV channel} - We are the first to demonstrate a method that extracts a targeted POI from an encrypted FPV channel and can be used to distinguish between the legitimate use of a drone that does not invade a subject's privacy and illegitimate use; a difference that depends on the angle of the video camera and not on the location of the drone, as demonstrated in Figure \ref{fig:RSSI}.

\item \textbf{External Interception Model} - Other studies \cite{saponas2007devices,liu2010video,203850} that aimed to classify video streams sent from a VOD supplier (e.g., Netflix, YouTube, etc.) to a client over the Internet used traffic that was captured from the client's network. This setup raises two problems: (1) in real life, in order to capture traffic from a targeted network, a computer inside the targeted network must be infected with a malware, and (2) most of these studies claim that their model works using external network interception. There are major challenges that arise when using an external interception model (e.g., packet loss), and their models' robustness to such challenges is unclear. Our study presents an external interception model using a radio frequency (RF) scanner that was empirically evaluated outside a lab setup and does not require any network infection.


\item \textbf{Effective for Encrypted Traffic} - Our methods work by analyzing encrypted traffic without any prior knowledge regarding the cryptography algorithm that is being used. We only use the length of the cryptogram which can be extracted from the second layer of the OSI model instead of its higher levels which were used in other studies \cite{saponas2007devices,liu2010video} to classify video streams.



\end{itemize}

The rest of this paper is structured as follows: in Section \ref{Sec:Background} we discuss various protocols and coding algorithms for video streams. In Section \ref{Sec:Related Work} we review related works from two main areas: information leakage from video streams and known methods of detecting nearby drones. In Section \ref{label:Adversary Model} we present (1) an adversarial model of an attacker that performs a privacy invasion attack, and (2) an external interception model for the detection of a captured POI. In Section \ref{section:Analysis} we investigate the influence of changing pixels on the transmitted traffic (in a lab setup), and in Section \ref{sec:Conclusions} we discuss legal solutions for privacy invasion attacks. 

\section{Background}
\label{Sec:Background}

Modern drones provide video piloting capabilities (FPV channel), in which a live video stream is sent from the drone to the pilot (operator) on the ground so the pilot is flying the drone as if he/she was onboard instead of looking at the drone from the pilot’s actual ground position. It allows an operator to control a drone using a remote controller. There are three types of technologies dominating the FPV market: Wi-Fi FPV and analog FPV. Wi-Fi FPV is, by far, the most popular method used to include FPV in budget RC drones because: (1) any Android/iOS smartphone (or tablet) on the market can be used to operate the drone, and (2) the additional hardware required includes only a Wi-Fi FPV transmitter (that is connected to the camera of the drone), instead of the dedicated radio transmitter and dedicated controller required by other FPV types (e.g., 5.8GHz Analog FPV and 2.4GHz Analog FPV). Almost every FPV-enabled drone selling for less than \$100 uses Wi-Fi FPV \cite{WiFi-FPV-vs-5.8GHz-FPV-vs-2.4GHz-FPV-Ultimate-Guide}; in addition, there are dozens of kinds of Wi-Fi FPV drones available for purchase \cite{Wifi-FPV-Drones,8-smartphone-controlled-drones,8-DRONES-THAN-CAN-BE-CONTROLLED-BY-A-SMARTPHONE} and their price varies from \$30 for a simple drone up to hundreds of dollars for a professional drone (DJI Mavic, DJI Spark, Parrot Bebop 2) with HD resolutions (up to 4K) and extended functionality (e.g., follow me, return home, etc.).

Wi-Fi communication between the controller and the drone is sent over a secured access point that is opened by either the drone or the controller (both parties are connected to the access point). Using dedicated hardware (e.g., a controller with a Wi-Fi signal range extender), current drone models provide operators the ability to control a drone using FPV from a distance of a few kilometers over Wi-Fi channels \cite{Spark-Remote-Controller,PARROT-SKYCONTROLLER, Phantom-range-extenders}.


The video that is captured by the drone camera is streamed to its controller using real-time end-to-end media streaming protocols (RTP). The RTP standard describes two sub-protocols: RTP (Real-Time Transport Protocol) and RTCP (RTP Control Protocol). \textbf{RTP} is used to transport media, and the majority of its implementations are built on the user datagram protocol (UDP), since real-time multimedia streaming applications require timely delivery of information and can often tolerate some packet loss to achieve this goal. \textbf{RTCP} transports statistics for a media connection and information such as the transmitted octet and packet counts, packet loss, packet delay variation, and round-trip delay time. RTP and RTCP each have a secured version \textbf{SRTP} and \textbf{SRTCP} which are intended to provide encryption, message authentication and integrity, and replay protection. \textbf{RTSP} is a network control protocol designed for use in entertainment and communication systems to control streaming media servers. The protocol is used for establishing and controlling media sessions between end points, and it supports a transmission of commands, such as play, record, and pause.

\subsection{Video Coding Algorithms}
\label{subsection:Video Coding Algorithms}

Video encoding \cite{wiegand2003overview, ostermann2004video, jack2011video} begins with a raw image captured from a camera. The camera converts analog signals generated from striking photons into a digital image format. Video is simply a series of such images generally captured five to 120 times per second (referred to as frames per second or FPS). The stream of raw digital data is then processed by a video encoder in order to decrease the amount of traffic that is required to transmit a video stream. Video encoders use two techniques to compress a video: intra-frame coding (spatial compression) and inter-frame coding (temporal compression).

\textbf{Intra-frame coding} creates an \textbf{I-Frame}, a time periodic reference frame that is strictly intra-coded. The receiver decodes an I-frame without additional information. Intra-frame prediction exploits spatial redundancy, i.e., correlation among pixels within one frame, by calculating prediction values through extrapolation from already coded pixels for effective delta coding.
The intra-coding process contains the following stages \cite{mitrovicvideo,jack2011video}:

\begin{enumerate}
\item Color conversion and chroma sub-sampling - The human eye has a lower sensitivity to color information than to dark-bright contrasts. First a conversion from RGB color space into YUV color components (e.g., YCbCr) is applied, and then, some of the chrominance information of the image is removed. This is a lossy stage.
\item Partition - The actual frame is divided into non overlapping macroblocks.
\item Transformation  - A block is represented in the frequency domain.
\item Quantization - This process is applied to the block to remove the insignificant part (high frequencies) and results in a compressed block with a smaller amount of information. This is a lossy stage.
\item Entropy coding - Compression algorithms are used to represent the data by mapping frequently occurring patterns with a few bits and rarely occurring patterns with many bits (e.g., using Huffman coding).

\end{enumerate}

Over the years various optimizations have been introduced for each of the stages, including: (1) dynamic partitioning techniques, (2) novel prediction algorithms and varying the amount of reference frames, (3) different domain transformations, and (4) quantization methods. These optimizations boost the transmission rate from 1.5 Mbps (MPEG-1) to 150 Mbps (MPEG-4).

\textbf{Inter-frame coding} exploits temporal redundancy by using a buffer of neighboring frames that contains the last M number of frames and creates a delta frame. A delta frame is a description of a frame as a delta of another frame in the buffer. The receiver decodes a delta frame using a received reference frame. There are two major types of delta frames: P-Frame and B-Frame. \textbf{P-Frames} can use data from previous frames to decompress and are more compressible than I-Frames. \textbf{B-Frames} can use both previous and upcoming frames for data reference to get the greatest amount of data compression. The process of generating a delta frame consists of the following stages: 
\begin{enumerate}
\item Partition - dividing the actual frame into nonoverlapping macroblocks.
\item Reference block matching - finding a similar block in another frame. 
\item Motion vector extraction - extracting the difference between the two blocks by calculating the prediction error. 
\end{enumerate}

\begin{figure}
\centering
\includegraphics[width=0.35\textwidth]{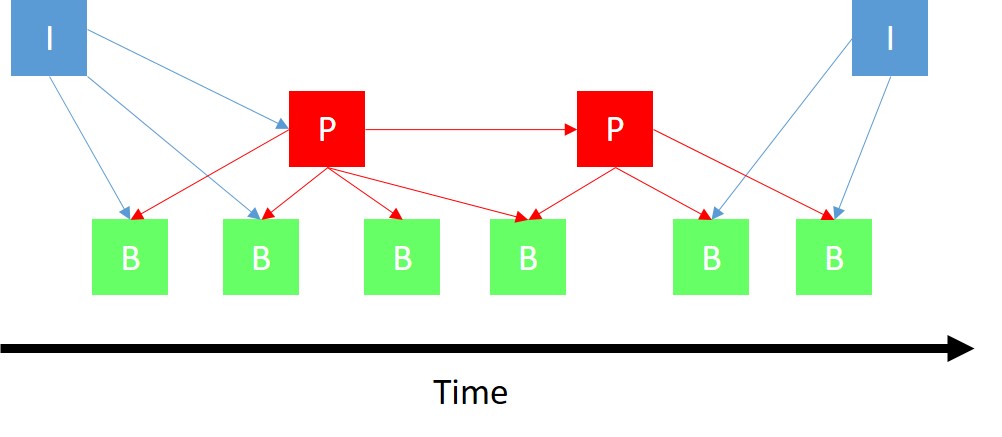}
\caption{GOP structure - I,B, and P-frames}
\label{fig:gop}
\end{figure}

The order in which I, B, and P-Frames are arranged is specified by a GOP (group of pictures) structure. A GOP is a collection of successive pictures within a coded video stream. It usually consists of two I-Frames, one at the beginning and one at the end. In the middle of the GOP structure, P and B-Frames are ordered periodically. An example of a GOP structure, with I, P, and B-Frames, can be seen in Figure \ref{fig:gop}. Occasionally B-Frames are not used in real-time streaming due to delays.

Intra-framing and inter-framing techniques were integrated into the MPEG-1 standard in the 1990s. Naturally, integrating these techniques into the protocol creates a variable bitrate (VBR) in the transmission of a video which is influenced by changes between frames and the content of the frame itself. A frame that can be represented as a set of prediction blocks of a similar neighboring frame (that has already been captured and transmitted) requires a smaller amount of data to be represented. The same thing is also true for video streams with a lot of redundancy in their frames. On the other hand, a frame with less similarity to other neighboring frames (e.g., as a result of the movement of several objects) necessitates that a larger amount of data be represented as a set of prediction blocks of other frames. The same thing is also true for a frame with less redundancy. Even if the video stream is encrypted at the transport layer (e.g., using TLS), the sizes of the packets and times of arrival are visible to anyone watching the network. In terms of cyber security such coupling between the captured stream and its cryptogram series can be used to extract meaningful information, as described below in Section \ref{Sec:Related Work}.

\section{Related Work}
\label{Sec:Related Work}

\begin{figure*}
\centering
\includegraphics[width=1.0\textwidth]{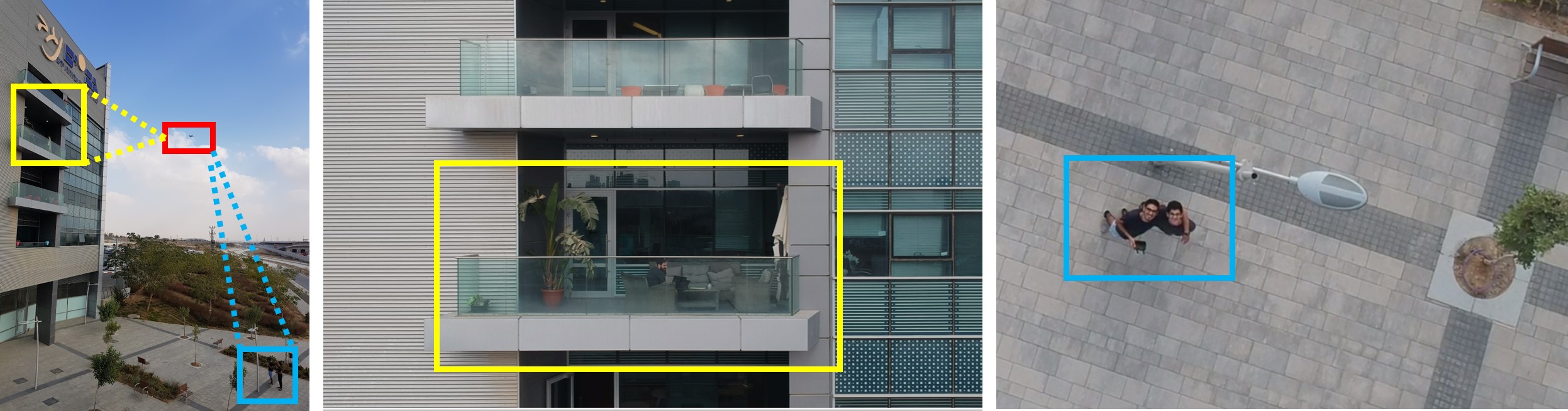}
\caption{From left to right: (a) a drone boxed in red, two people boxed in blue, and a window of an organization boxed in yellow, (b) illegitimate use of the drone camera - filming the organization, (c) a legitimate use for selfie purposes.} 
\label{fig:RSSI}
\end{figure*} 

In this section we describe: (1) methods that exploit information leakage of an encrypted video stream to extract insights about the stream, and (2) methods for nearby drone detection. In the area of \textbf{video hosting services}, several studies exploited video stream information leakage to classify a video stream sent from a video hosting service (e.g., YouTube, Netflix, etc.) using Dynamic Adaptive Streaming over HTTP (DASH) protocols (a.k.a. MPEG-DASH). This attack model relies on two steps: (A) building a database of reference traces of video streams, and (B) classifying a query trace of an intercepted video stream by matching it to the database. \textit{Saponas et al.} \cite{saponas2007devices} analyzed Slingbox's encrypted streams sent over wired and wireless connections to a client installed on a computer and managed to achieve a 89\% accuracy in classifying 26 different movies by analyzing 40 minutes of the stream's bitrate. \textit{Schuster et al.} \cite{203850} classified video streams sent from Netflix, Amazon, YouTube, and Vimeo by analyzing burst patterns using convolutional neural networks. \textit{Reed et al.} \cite{reed2016leaky, reed2017identifying} classified Netflix's video streams and reached accuracy over 90\% by analyzing only eight minutes of the stream. In an area similar to video hosting services, \textit{Liu et al.} \cite{liu2010video} constructed robust video signatures using wavelet based analysis by analyzing traffic sent over the RTP of an \textbf{IPTV}. In the area of \textbf{VoIP}, \textit{Wright et al.} showed that VBR leakage in encrypted VoIP communication can be used for the detection of the speaker's language \cite{wright2007language} and phrases \cite{wright2008spot}. \textit{White et al.} \cite{white2011phonotactic} extended this approach to extract conversation transcripts. \textit{Wampler et al.} \cite{7417767} analyzed packets' average inter-arrival time, size, and bandwidth, and the number of received packets in a window, in order to extract hand movement and ambient light changes from \textbf{IP camera} traffic sent over RTP.

In terms of the attack model, the described studies did not conduct their experiments using an external RF scanner (e.g., NIC in monitor mode), so their attack model requires a malware installed on the targeted network/computer in order to detect the video streams. 

In the area of \textbf{drone detection}, various methods were introduced over the last few years to detect a nearby drone. Radar is a traditional method of detecting drones, however the detection of small consumer drones requires expensive high-frequency radar systems \cite{eshel2013mobile}. Several studies suggested computer vision techniques to detect a drone by using a camera to analyze motion cues  \cite{rozantsev2015flying,busset2015detection}. However, these methods suffer from false positive detections due to: (1) the increasing number of drone models, and (2) the similarities between the movements of drones and birds \cite{busset2015detection}. In order to distinguish between birds and drones, several approaches analyzed  the noise of the rotors captured by microphones \cite{case2008low,busset2015detection}. However, very expensive equipment is required in order to address the challenges arising from the ambient noise and the distance between the drone and the microphone \cite{case2008low}. A hybrid method that combines all of the methods discussed in this section was suggested by \cite{vasquez2008multisensor} in order to improve the accuracy of detection, however such a method is very expensive to deploy. Two other studies proposed a method to detect a consumer/civilian drone controlled using Wi-Fi signals. The first method \cite{peacock2013towards} analyzes the protocol's signatures of the Wi-Fi connection between the drone and its controller. The second method \cite{birnbach2017wi} analyzes the received signal strength (RSS) using a RF scanner (e.g., Wi-Fi receiver).

None of the described methods for drone detection is able to determine whether the drone was used to invade privacy (by video recording the subject/target). More specifically, they are unable to understand what exactly is being recorded by the drone. In crowded areas, the difference between legitimate and illegitimate use is based on the angle of the drone's camera. Figure \ref{fig:RSSI} presents legitimate and illegitimate uses of a drone. All of the described methods \cite{eshel2013mobile, rozantsev2015flying, busset2015detection, case2008low, vasquez2008multisensor, peacock2013towards, birnbach2017wi} fail to distinguish between the act of taking a selfie and a privacy invasion attack. In contrast, our method does not have these abovementioned weakness. In this research we demonstrate methods for: (1) determining exactly what is being recorded, and (2) providing a subject with proof that he/she was under surveillance.  

\begin{table}[tpb]
	\centering
    \caption{Information leakage from VBR - related work}
	\resizebox{1.0\columnwidth}{!}{%
        \begin{tabular}{|l|l|l|l|l|l|}
        \hline
        \begin{tabular}[c]{@{}l@{}}		   \end{tabular}

\textbf{Transmitter}
			& \textbf{Purpose}
            & \textbf{Publication}
            & \begin{tabular}[c]{@{}l@{}} \textbf{Required}\\ \textbf{.o'
            } \end{tabular}
			& \begin{tabular}[c]{@{}l@{}}			\textbf{Analyzed}\\ \textbf{Protocols} \end{tabular}
            & \begin{tabular}[c]{@{}l@{}}			\textbf{Interception} \end{tabular}\\\hline
            \begin{tabular}[c]{@{}l@{}}		Video Hosting\\Services (Netflix,\\YouTube, etc.)\end{tabular} 
			& \begin{tabular}[c]{@{}l@{}}		Classify\\Video-stream \end{tabular}
            & \begin{tabular}[c]{@{}l@{}}		
            \cite{saponas2007devices} - USENIX 2007\\ \cite{liu2010video} - \\ \cite{203850} - USENIX 2017\\ \cite{reed2017identifying} - CODASPY 2017\\ \cite{reed2016leaky} - CCNC 2016\end{tabular}
            &  \begin{tabular}[c]{@{}l@{}}Minutes\end{tabular} 
			& DASH	 
            & Internal\\\hline
            \begin{tabular}[c]{@{}l@{}}		IPTV\end{tabular} 
			& \begin{tabular}[c]{@{}l@{}}		Classify\\Video-stream\end{tabular}
            & \cite{liu2008wavelet} - GLOBECOM 2008
            &  Minutes
			& RTP	 
            & Internal\\\hline
            IP Camera
			& \begin{tabular}[c]{@{}l@{}}		Lights on/off,\\Hand movement \end{tabular} 
			& \cite{7417767} - GLOBECOM 2015
            & Immediate
            & RTP
            & Internal\\\hline            
            PC
			& \begin{tabular}[c]{@{}l@{}}		Language extraction\cite{wright2007language}\\Phrase detection\cite{wright2008spot}\\Transcripts\cite{white2011phonotactic}\\ \end{tabular}             			
            & \begin{tabular}[c]{@{}l@{}}\cite{wright2007language} - USENIX 2007\\ \cite{wright2008spot} - S\&P 2008\\ \cite{white2011phonotactic} - S\&P 2011 \end{tabular}             			
            & -            
            & VoIP
            & Internal\\\hline
            \rowcolor{yellow}
            {\bfseries \textbf{Drone}}
			& \begin{tabular}[c]{@{}l@{}}\textbf{Detecting Streamed POI}  \end{tabular}
            & -
			& \textbf{10 seconds}
            & {\bfseries \textbf{RTP}}            
            & {\bfseries \textbf{External}}\\\hline
		\end{tabular}
        
	}
~\label{tab:related_work}
\end{table}

\section{Adversary Model \& Proposed Detection Scheme}
\label{label:Adversary Model}

We consider an adversary operator that uses a drone to film a target (subject or organization) for:
\begin{enumerate}
\item Self-entertainment - the attacker considers a privacy invasion attack as a form of entertainment and performs the attack to satisfy his/her curiosity.
\item Malicious purpose - the attacker uses the drone's video camera to collect information about the target for malicious purposes. For example, in cases in which the target is an organization, the malicious purpose can be to break into an organization (the drone can be used to count the number of subjects that leave the building). Another malicious purpose is using the drone to disable a secret facility (in which the captured video is used to map the organizational assets). In cases in which the target is a subject, the purpose can be to understand whether the subject is cheating on his/her spouse by using the drone's video camera to spy on the subject (as was shown in \cite{cheating}).
\end{enumerate}

\begin{figure}
\centering
\includegraphics[width=0.50\textwidth]{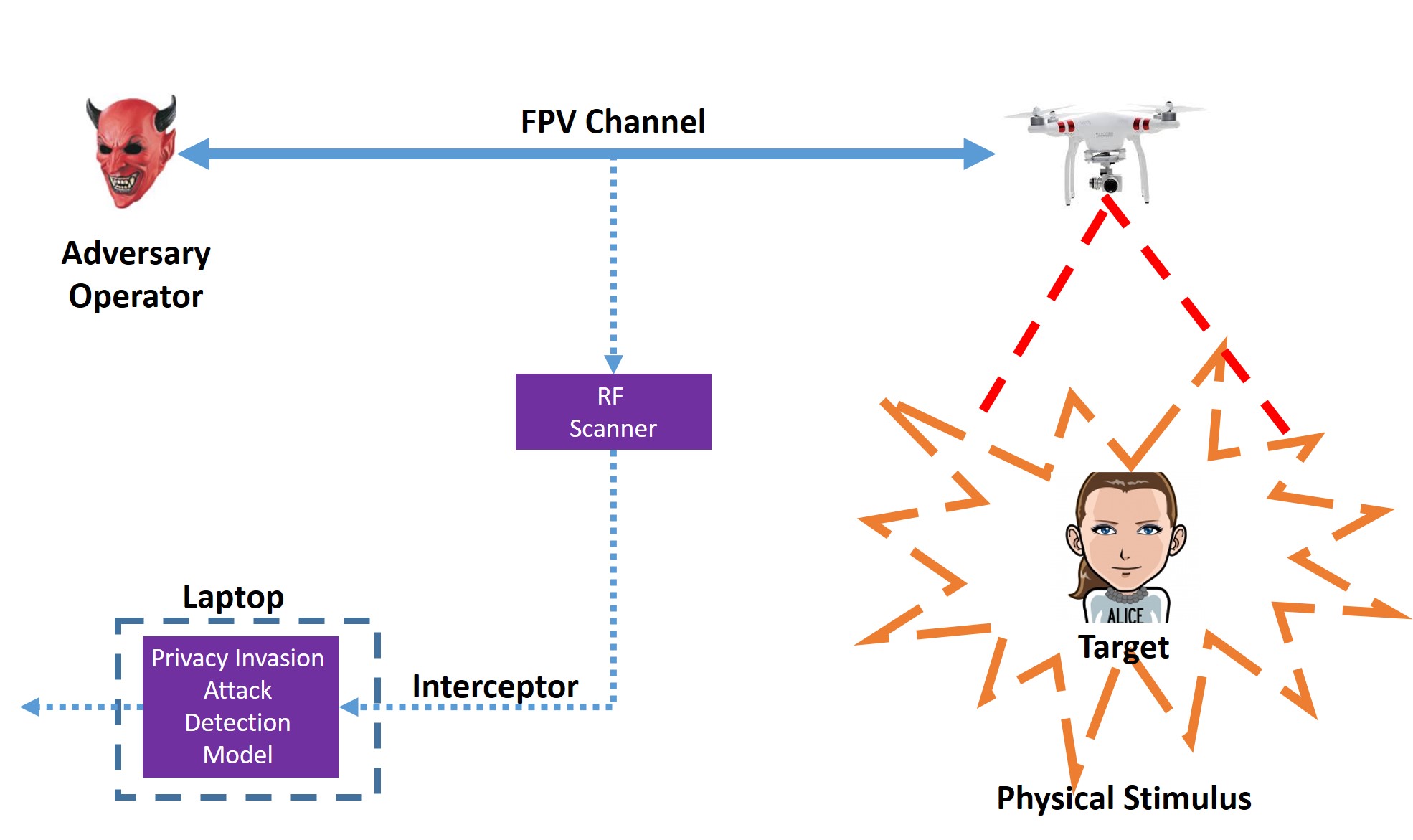}
\caption{Proposed target detection scheme.}
\label{fig:model}
\end{figure}

The interceptor's goal is to determine whether a target is being captured by a drone's video camera. We assume that an interceptor has detected the presence of a drone nearby (using one of the known methods for drone detection \cite{eshel2013mobile, rozantsev2015flying, busset2015detection, case2008low, vasquez2008multisensor, peacock2013towards, birnbach2017wi}) or by analyzing suspicious access points. In addition we assume that the interceptor owns an RF scanner (e.g., an NIC) that is connected to a computer with an adequate antenna that captures the traffic being sent from the drone to the controller. The interceptor initiates a physical stimulus aimed at the captured target in a random pattern and analyzes the intercepted traffic in a detection model. Figure \ref{fig:model} presents the proposed target detection scheme and the parties involved. 

\subsection{Detection Model}

 Algorithm \ref{algorithm-lights} presents the target detection model. It receives as input: (1) the intercepted Wi-Fi stream (that was captured by the NIC), (2) a watermarking pattern (binary sequence) that was modulated by the physical stimulus, and (3) a window for each bit that was modulated. In addition, the algorithm receives the (4) start and (5) end time of the watermarked pattern (in epoch representation). First, the intercepted Wi-Fi traffic is converted to a bitrate array (line 3). A $stableInterval$ is extracted for the following time period: four seconds before the first physical stimulus begins until the next physical stimulus starts (line 5). A $stimulusInterval$ is extracted for the following time period: for four seconds from the time at which the first physical stimulus begins (line 6). A $stableBitrate$ and $stimulusBitrate$ are calculated by averaging $stableInterval$ and $stimulusInterval$ intervals (lines 8-9). A $cutoff$ is calculated as the middle between the  $stableInterval$ and the $stimulusBitrate$ (line 10). Each bit interval is extracted from $bitrateArray$,  classified as 0/1 bits and concatenated to $extractedPattern$  (lines 11-17). Finally, a Boolean result is returned by comparing the $watermarkedPattern$ to $extractedPattern$ patterns(line 18). 

\begin{algorithm}
\caption{Privacy Invasion Attack Detection}\label{algorithm-lights}
\hspace*{\algorithmicindent} \textbf{Input:} \\ 
\hspace*{\algorithmicindent} 1) intercepted-WiFI-Stream // Intercepted by the NIC\\
\hspace*{\algorithmicindent} 2) watermarkingPattern // binary sequence (e.g., 101..01)\\ 
\hspace*{\algorithmicindent} 3) window // milliseconds for single bit modulating\\
\hspace*{\algorithmicindent} 4) beginPatternTime // begin time of pattern (epoch)\\ 
\hspace*{\algorithmicindent} 5) endPatternTime // end time of pattern (epoch)\\ 
\hspace*{\algorithmicindent} \textbf{Output:} \\
 \hspace*{\algorithmicindent}  Boolean result

\begin{center}
\line(1,0){250}
\end{center}

\begin{algorithmic}[1]
\Procedure{\textbf{underDetection?}}{}
\State $\textit{extractedPattern}  \gets \text{""}$
\State $\textit{bitrateArray} = \textit{extractBitrateArray(intercepted-WiFI-Stream)}$
\State $\textit{stableBeginTime} = \textit{beginPatternTime - 4000}$
\State $\textit{stableInterval} = \textit{bitrateArray.subarray(stableBeginTime, beginPatternTime)}$
\State $\textit{endStimulusTime} = \textit{beginPatternTime + 4000}$
\State $\textit{stimulusInterval} = \textit{bitrateArray.subarray(beginStimulusTime,endStimulusTime)}$
\State $\textit{stableBitrate}  = \textit{average(stableInterval)}$
\State $\textit{stimulusBitrate}  =\text{average(stimulusInterval)}$
\State $cutoff = (stumulusBitrate + stableBitrate)/2$
\For{$\textit{(i = beginPatternTime; i < endPatternTime; i = i+window)}$}	
\State $\textit{interval}=\textit{bitrateArray.subArray(i,i+window)}$
\State $\textit{avg} = \textit{average(interval)}$
    \If{avg > cutoff}
    \State $\textit{result} = \textit{result + "1"}$
    
\Else     \State $\textit{result} = \textit{result + "0"}$
\EndIf

\EndFor

\State return (watermarkingPattern == extractedPattern)
\EndProcedure

\end{algorithmic}
\end{algorithm}

\subsection{Intercepting FPV Channels}

As was discussed in Section \ref{Sec:Background}, many commercial drones provide FPV capabilities over Wi-Fi channels. In this case, the drone/controller exposes a secured access point that both parties connect to using authentication. The video stream is sent over Wi-Fi communication and can be intercepted using an NIC (in monitoring mode). An antenna can be used by an interceptor in order to extend the interception range. DJI Mavic, DJI Spark, and Parrot Bebop 2 drones use Wi-Fi Protected Access II (WPA2) protocols to secure their networks. The access points of FPV channels can be detected by changing the NIC mode of a laptop to monitoring mode (or using a software defined radio instead) and using dedicated tools (such as airmon, inSSIDer, etc.), that can even detect hidden networks, in order to find suspicious access points. After identifying suspicious access points, a specific access point can be found by searching for known BSSIDs or MAC IDs of drones. In situations in which the BSSIDs or MAC IDs have been changed, the interceptor can use a method to detect the type of drone by performing forensic analysis of the access point communication as was suggested by \textit{Peacock et al.} \cite{peacock2013towards}. Another option is to analyze each of the access points within range of the target in the detection model. 

\section{Influence of Physical Stimulus}
\label{section:Analysis}

In this section we investigate the influence of physical stimuli (that change pixels in the streamed video) on the transmitted traffic in a lab setup. All of the methods described in this section make use of a simple principle that changes in the number of pixels from a frame to a consecutive frame requires data to encode, therefore changing a large number of pixels results in more data to encode and causes the FPV's bitrate to increase (intra-frame coding). We show how the act of flickering an object within a streamed video changes the FPV's bitrate and test the influence of dividing the flickering object and changing its flickering colors.

\subsection{Lab Experiments}
\label{subsection:Active Approach - Lab Experiments}

\begin{figure*}
\centering
\includegraphics[width=1.0\textwidth]{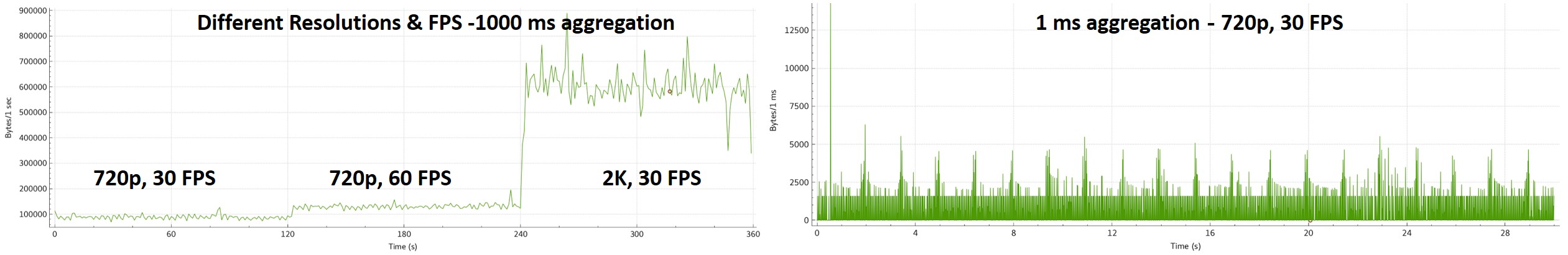}
\caption{Bitrate of captured Wi-Fi signals of the white wall at different resolutions}
\label{fig:static}
\end{figure*}

In the preliminary lab experiments described below we assess the influence of various changes to the pixels on the traffic using a Mavic Pro \cite{Mavic-Pro} consumer drone. The drone was configured to transmit video at a rate of 24 frames per second (FPS), its default configuration. We used a laptop (Dell Latitude 7480) that runs Kali Linux with a standard NIC (Intel Dual-Band Wireless-AC 8265 Wi-Fi \cite{Intel-Dual-Band-Wireless-AC-8265}) for interception. We enabled the monitor mode on the NIC using airmon-ng \cite{Airmon-ng} and intercepted the encrypted video traffic of the Mavic's AP. The Mavic's AP uses 802.11n to transfer the data between the connected parties. From the external interception perspective, we were able to extract only the second layer (data link layer) meta-data which includes the following: BSSID, source MAC address, destination MAC address, and packet length. The payload of the packet is encrypted. 

We started by analyzing the Mavic's traffic when the captured video is steady by placing the Mavic in front of a white wall. Figure \ref{fig:static}a shows the bitrate of the traffic that was transmitted from the drone for a period of 240 seconds and intercepted by a laptop's NIC (in monitoring mode) at 1000 ms aggregation at three different resolutions and rates (720p 30 FPS, 720p 60 FPS, and 2K 30 FPS). As can be seen from the results in Figure \ref{fig:static}, the bitrate is fairly stable for each resolution over time, however higher resolutions generate higher bitrates. 

\begin{figure}
\centering
\includegraphics[width=0.50\textwidth]{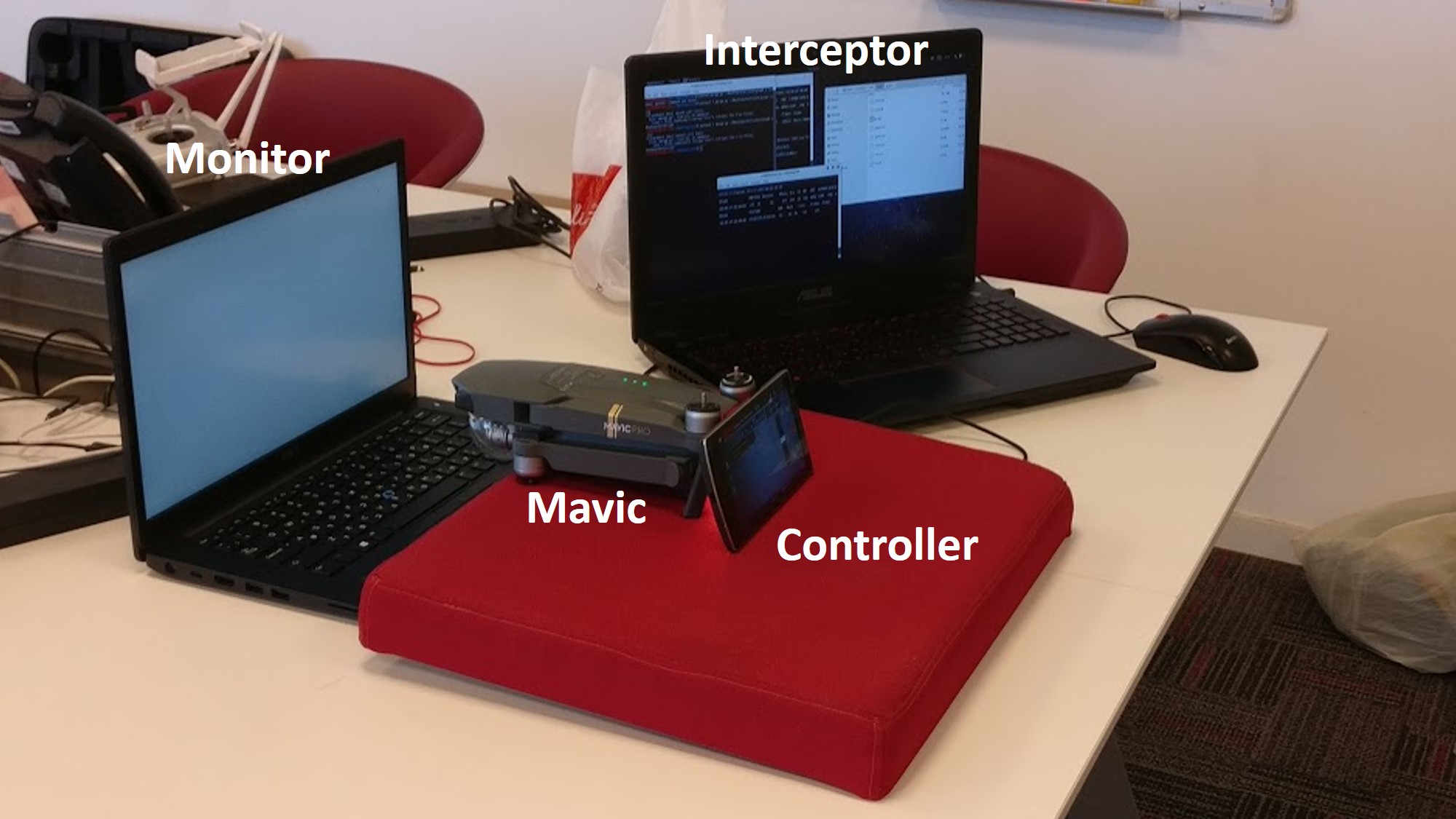}
\caption{Lab Setup - The DJI Mavic is placed in front of a laptop monitor. A second laptop is  used to intercept the traffic (using its NIC in monitoring mode).}
\label{fig:lab_setup}
\end{figure}

In the rest of the experiments described in this section, we placed the Mavic in front of a laptop in order to expose the drone to specific images/objects on the monitor. The experimental setup is presented in Figure \ref{fig:lab_setup}.

First, we investigated the effect of changing the percentage of captured pixels on the traffic. We programed a Python code to present a flickering rectangle on the screen in the middle of the monitor. We tested the effect of various rectangle sizes on the traffic that was sent from the drone to the controller using external interception (the interception laptop was not connected to an access point; its NIC was on monitor mode). The rectangle flickered from white to black, and vice versa, on a white background for 40 seconds. 

\begin{table}[]
	\centering
    \caption{Influence of changing the amount of pixels on the traffic}
	\resizebox{1.0\columnwidth}{!}{%
       \begin{tabular}{|l|l|l|l|l|l|l|l|l|l|}
    \toprule
    \multicolumn{1}{|p{7.315em}|}{Percentage of \newline{}changing pixels} & 0\%   & 1.2\% & 2.50\% & 5\%   & 10\%  & 25\%  & 50\%  & 75\%  & 100\% \\
    \midrule
    Bitrate (KB) & 120   & 130   & 135   & 161       & 170       & 230     & 260     & 290       & 320       \\
    \midrule
    Delta Bitrate (KB) & 0     & 10    & 15    & 41       & 50       & 110       & 140       & 170       & 200       \\
    \midrule
    Delta Bitrate (\%) & 0\%   & 8\%   & 13\%  & 34\%  & 42\%  & 92\%  & 117\% & 142\% & 167\% \\
    \bottomrule
    \end{tabular}%
	}
~\label{tab:sizes}
\end{table}

As can be seen from the results presented in Table \ref{tab:sizes}, there is a strong connection between the percentage of pixels changed and the volume of traffic that was sent from the drone. This phenomenon occurs because a larger amount of changing pixels results in a larger amount of changing macroblocks. A larger amount of changing macroblocks means that the encoder must use more data to encode the delta frames; this in turn increases the amount of traffic. In addition, as the results show, very small changes (< 2.5\%) are effectively absorbed and merged with the background noise.

Next, we aimed to determine the effect of separating the pixels and dividing them across several objects (rather than centralizing them on one object) on the amount of traffic generated (given a fixed number of changed pixels). In this series of experiments we fixed the amount of changing pixels but presented a different number of rectangles, dividing the fixed number of pixels to form smaller equal sized rectangles (2, 4, 8, 16, 32), and positioned the rectangles in different places on the monitor. As can be seen from the results presented in Table \ref{tab:division}, there is a strong connection between increasing the number of rectangles and an increase in the amount of traffic that is required to encode the change. We believe that this phenomenon can be explained as follows: dividing a single rectangle (which centralizes the fixed number of pixels) into smaller pieces (thereby dividing the fixed number of pixels) and separating them from each other on the monitor results in the intersection with more macroblocks that change compared to a centralized object of the same size. Therefore, this requires more data to encode and increases the amount of traffic.

\begin{table}[]
	\centering
    \caption{Influence of dividing an area into pieces on the traffic}
	\resizebox{1.0\columnwidth}{!}{%
    \begin{tabular}{|l|l|l|l|l|l|l|}
    \toprule
    Number of Pieces & 1     & 2     & 4     & 8     & 16    & 32 \\
    \midrule
    Bitrate (KB) & 250   & 260   & 275   & 300   & 325   & 340 \\
    \midrule
    Bitrate Delta (KB) & 0     & 10    & 25    & 50    & 75    & 90 \\
    \midrule
    Bitrate Delta (\%) & 0.00\% & 4.00\% & 10.0\% & 20.0\% & 30.0\% & 36.00\% \\
    \bottomrule
    \end{tabular}%
	}
 ~\label{tab:division}
\end{table}

Finally, we assessed whether the objects' position on the monitor affects the traffic. In order to do this, we conducted an experiment in which we flickered a rectangle that is one fourth the size of the screen in four different places on the monitor: top right, top left, bottom left, and bottom right. As can be seen from the results presented in Table \ref{tab:locations}, each of the flickering rectangles had the same effect on the traffic. Therefore, we believe that when the objects' size remains fixed, the location on the monitor has no effect, since the same number of changing macroblocks is involved. 

\begin{table}[]
	\centering
    \caption{Influence of location of an object on the traffic}
	\resizebox{1.0\columnwidth}{!}{%
    \begin{tabular}{|l|l|l|l|l|l|}
    \toprule
          & White Screen & Top Left & Top Right & Bottom Left & Bottom Left \\
    \midrule
    Bitrate (KB) & 120   & 195   & 195   & 195   & 195 \\
    \midrule
    Bitrate Delta (KB) & 0     & 75    & 75    & 75    & 75 \\
    \midrule
    Bitrate Delta (\%) & 0.00\% & 62.500\% & 62.500\% & 62.500\% & 62.500\% \\
    \bottomrule
    \end{tabular}%
	}
 ~\label{tab:locations}
\end{table}

From this set of experiments we were able to conclude that (1) the larger the number of changed pixels, the greater the influence on traffic (larger number of changing macroblocks), and (2) the influence is even greater if the pixels are not clustered together (intersection with a larger number of macroblocks).

After investigating the effect of the number of pixels changed, the objects' location on the monitor, and the difference between keeping the pixels centralized vs. dividing them, we moved on to assess the effect of the object's color on the amount of traffic. We conducted an experiment in which we flickered different colored rectangles (black, green, blue, red, orange, yellow, pink, purple, and white) of the same size on the monitor.

\begin{table}[]
	\centering
    \caption{Influence of changing the color of an object on the traffic}
	\resizebox{1.0\columnwidth}{!}{%
        \begin{tabular}{|l|l|l|l|l|l|l|l|l|l|l|}
    \toprule
          & No flicker & Black & Red   & Green & Orange & Yellow & Gray  & Purple & Blue  & Pink \\
    \midrule
    \multicolumn{1}{|p{10.565em}|}{RGB Value} &       &       &       &       &       &       &       &       &       &  \\
    \midrule
    Bitrate (KB) & 100   & 325   & 325   & 325       & 325       & 325     & 325     & 325       & 325       & 325       \\
    \midrule
    Delta Bitrate (KB) & 0     & 225   & 225   & 225       & 225       & 225       & 225       & 225       & 225       & 225       \\
    \midrule
    Delta Bitrate (\%) & 0\%   & 225\% & 225\% & 225\% & 225\% & 225\% & 225\% & 225\% & 225\% & 225\% \\
    \bottomrule
    \end{tabular}%
	}
 ~\label{tab:colors}
\end{table}

As can be seen from results presented in Table \ref{tab:brightness}, each color caused the same effect on the amount of traffic that was sent from the drone. From this experiment we can conclude that no color significantly outperforms another.


Rather than using the RGB color space, video encoders use different color spaces to represent a picture including: YCbCr, YCoCg, etc. Video encoders transform a captured picture from an RGB color space to a luma value (denoted as Y) and two chroma values. The Y component can be stored with a high resolution or transmitted at a high bandwidth, and the two chroma components can be bandwidth-reduced, subsampled, compressed, or otherwise treated separately for improved system efficiency. Considering this information, we then tested the effect of different brightness levels of the same color on the traffic. To do so, we conducted an experiment in which we flickered two colors (green and blue) at five different brightness levels.

\begin{table}[]
	\centering
\caption{Influence of brightness level of an object on the traffic}
\resizebox{1.0\columnwidth}{!}{%
    \begin{tabular}{|l|l|l|l|l|l|l|}
    \toprule
    Brightness & No flicker & 0\%   & 20\%  & 40\%  & 60\%  & 80\% \\
    \midrule
    Bitrate (KB) & 100   & 300   & 300   & 310   & 320   & 350 \\
    \midrule
    Bitrate Delta (KB) & 0     & 200   & 200   & 210   & 220   & 250 \\
    \midrule
    Bitrate Delta (\%) & 0\%   & 200\% & 200\% & 210\% & 220\% & 250\% \\
    \bottomrule
    \end{tabular}%
	}
 ~\label{tab:brightness}
\end{table}

As can be seen from the results presented in Table \ref{tab:brightness}, increasing the level of brightness of the object increases the amount of traffic sent from the drone to the controller. The results obtained from the captured traffic were identical for blue and green colors. From these results, we concluded that brighter shades outperform darker shades of the same color.

In the next section we leverage our findings to detect privacy invasion attacks targeting a private house and a subject.




\section{Evaluation}
\label{section:Evaluation}
In this section we present the evaluation for target detection in two use cases: when the target is a private house and when the target is a subject.

\subsection{Detecting a Privacy Invasion Attack Against a Building}

In this subsection we demonstrate a method of securing a building from privacy invasion attacks by triggering a physical stimulus. As was shown in the lab experiments, flickering objects influence the amount of traffic that is required to encode the flickering objects. Taking this into consideration, we show how a smart film (a.k.a. smart glass) can be used as a means of triggering a physical stimulus in order to detect whether a building is being tracked by a drone. We purchased a smart film with an RF controller and attached it to a window of a private house that we wanted to secure. The smart film switches between two modes: transparent and white given a radio command sent from its controller. 

We wrote a simple Python program that uses a software defined radio (HackRF) to modulate a given signal using on-off keying (OOK) modulation. Each bit of the signal was modulated using a window of two seconds. For 1 bits we flickered the smart film, and for 0 bits we switched the smart film so it was transparent for the entire period of time. We randomly selected the sequence 111100001111111000000 as a signal to be modulated using the physical stimulus.

\begin{figure}
\centering
\includegraphics[width=1.0\columnwidth]{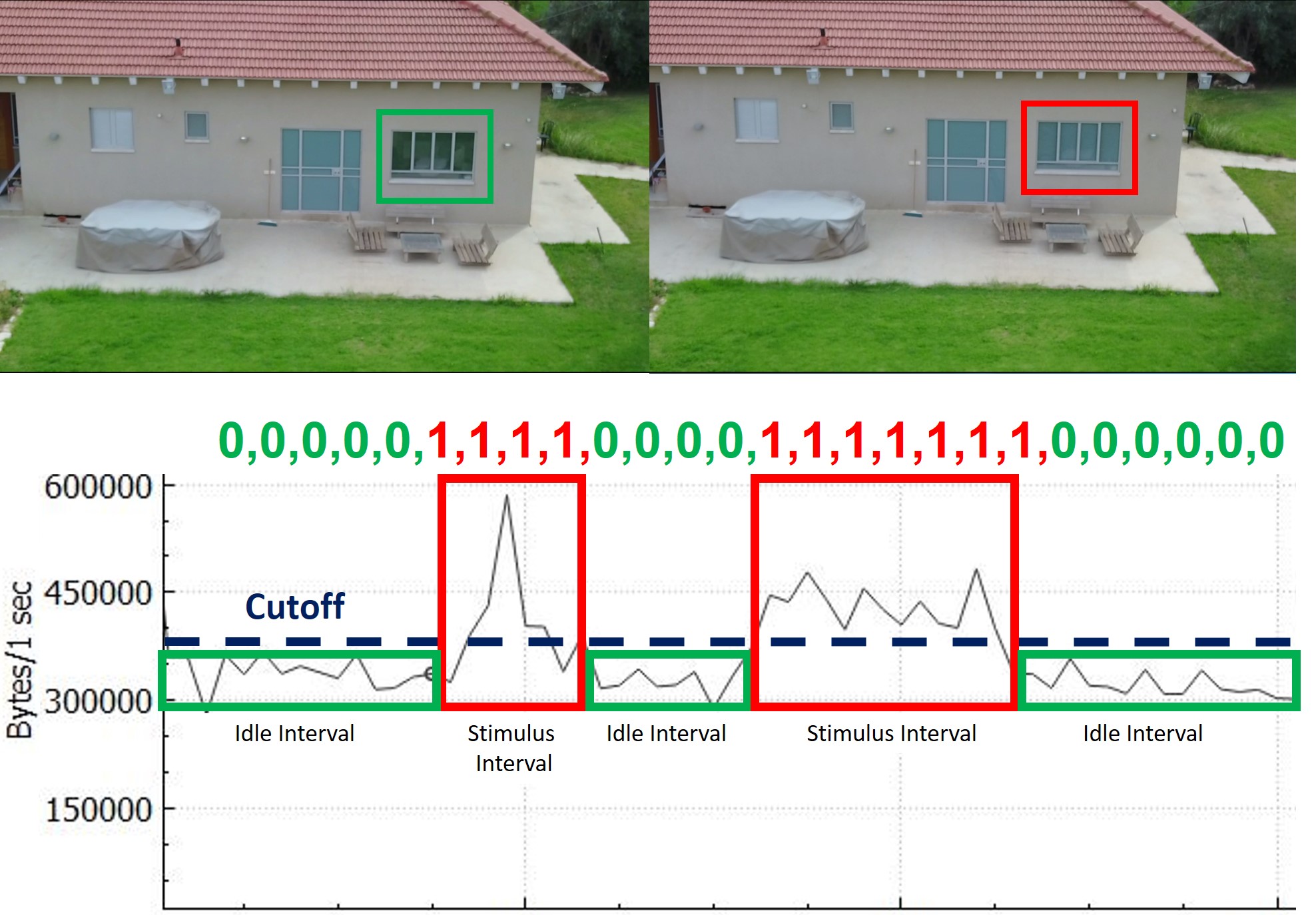}
\caption{A smart film controlled via a HackRF connected to a laptop.}
\label{fig:smart-film}
\end{figure}

In order to demonstrate an illegitimate use of the drone, we asked an operator to fly the DJI Mavic and film his neighbor's garden and private house from the operator's property. The default resolution and FPS were used (720p and 24 FPS). We purchased a parabolic antenna, TP-Link TL-ANT2424B 2.4GHz 24dBi Grid Parabolic, and connected it to a laptop to intercept the drone's outgoing traffic sent over the access point. We ran our Python code to create a physical stimulus using the smart film. Figure \ref{fig:smart-film} presents two snapshots that were taken from the streamed video and the results of applying Algorithm \ref{algorithm-lights} to the intercepted traffic. The peaks in the bitrate correlate to the time at which the smart film was flickered. The flicker that was used to modulate the 1 bits increased the bitrate up to 1.5-2 times, from an average of 300-350 KB to 450-570 KB. As can be seen from the intercepted traffic, the flicker that was produced using the smart film influenced the bitrate in a way that watermarked the bitrate according to the given pattern that was programmed in the Python code.

\begin{table}[]
	\centering
	\caption{FPR based on applying Algorithm \ref{algorithm-lights} on the same house without physical stimulus and another house} 
    \resizebox{0.7\columnwidth}{!}{%
    \begin{tabular}{|l|l|l|}
    \toprule
    \multicolumn{1}{|p{4.835em}|}{\textbf{Duration of \newline{} Stimulus\newline{}(in seconds)}} & \multicolumn{1}{p{6.75em}|}{\textbf{Target's house \newline{}without stimulus}} & \multicolumn{1}{p{7.125em}|}{\textbf{Neighbor's \newline{}house }} \\
    \midrule
    2     & 0.480 & 0.535 \\
    \midrule
    4     & 0.294 & 0.294 \\
    \midrule
    6     & 0.202 & 0.200 \\
    \midrule
    8     & 0.154 & 0.145 \\
    \midrule
    10    & 0.027 & 0.032 \\
    \midrule
    12    & 0.015 & 0.017 \\
    \midrule
    14    & 0.008 & 0.011 \\
    \midrule
    16    & 0.003 & 0.006 \\
    \midrule
    18    & 0.000 & 0.002 \\
    \midrule
    20    & 0.000 & 0.002 \\
    \midrule
    22    & 0.000 & 0.002 \\
    \midrule
    24    & 0.000 & 0.001 \\
    \midrule
    26    & 0.000 & 0.001 \\
    \midrule
    28    & 0.000 & 0.001 \\
    \midrule
    30    & 0.000 & 0.001 \\
    \midrule
    32    & 0.000 & 0.000 \\
    \midrule
    34    & 0.000 & 0.000 \\
    \midrule
    36    & 0.000 & 0.000 \\
    \bottomrule
    \end{tabular}%

	}
~\label{tab:yuval-fpr}
\end{table}

\begin{figure}
\centering
\includegraphics[width=1.0\columnwidth]{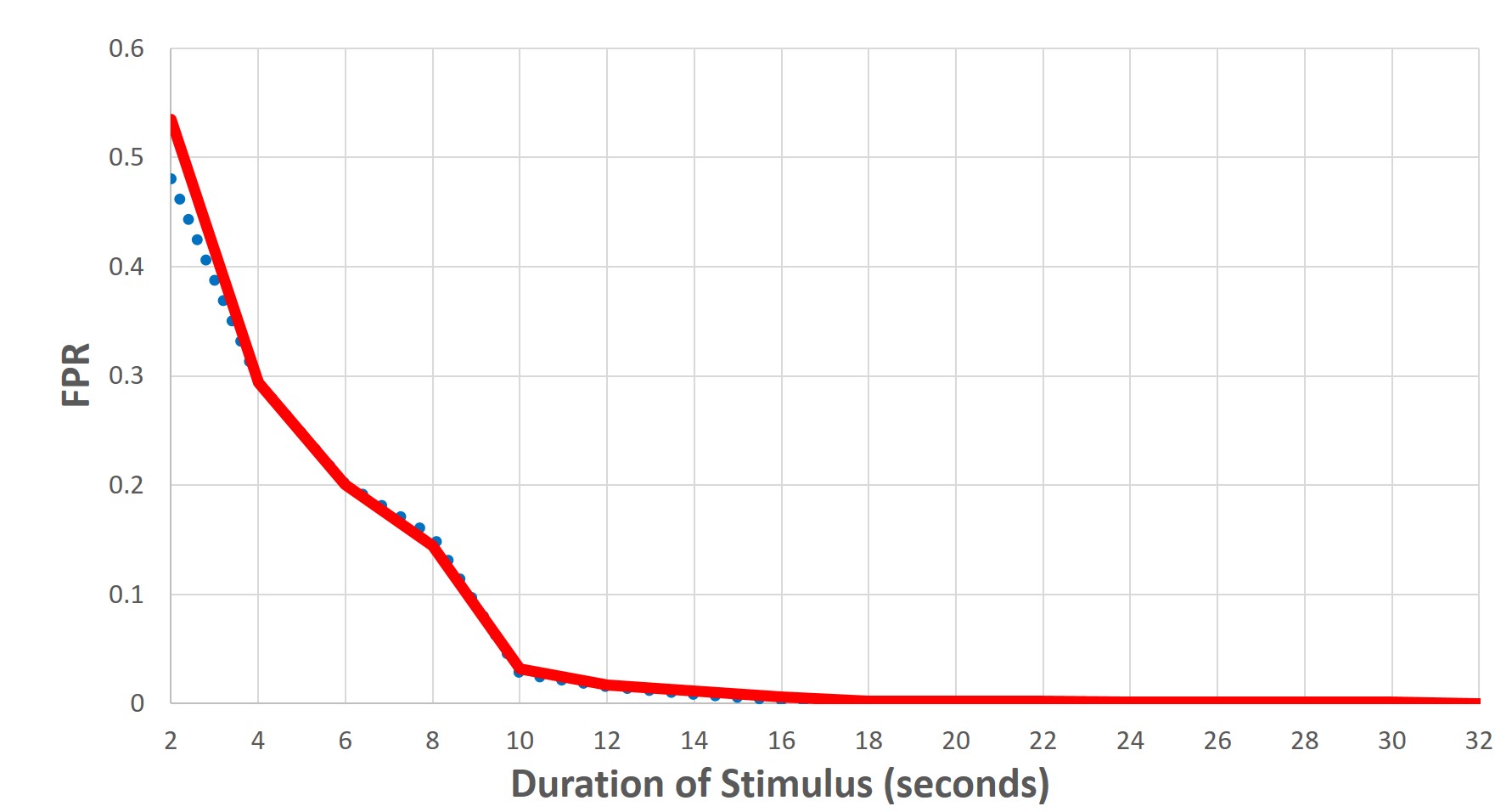}
\caption{A graph of the FPR as a function of the duration of the physical stimulus.}
\label{fig:yuval-graph}
\end{figure}

In order to prove that the physical stimulus is the cause of the traffic change, we conducted another experiment in which we streamed the same house for 20 minutes without initiating a physical stimulus. In addition, in order to prove that this effect could not be reproduced with another house (that is not the target), we streamed another house (the neighbor's house) for 20 minutes. In both cases we intercepted the traffic using the same experimental setup. We applied Algorithm \ref{algorithm-lights} to the intercepted traffic and calculated the false positive rate as a function of the duration of the physical stimulus for the original signal 111100001111111000000. As can be seen from the results that are presented in Table \ref{tab:yuval-fpr}, a pattern of 10 seconds is sufficient to exclude detection mistakes of filming another target (with an FPR of 0.032). In addition, a pattern of 10 seconds is sufficient to exclude mistakes of the same generated pattern without any physical stimulus that are coincidental and the  result of wind or other physical movement (with an FPR of 0.027).  Figure \ref{fig:yuval-graph} presents a FPR graph as a function of the duration of the physical stimulus.

\subsection{Detecting a Privacy Invasion Attack Against a Subject}

In this section we demonstrate a method to secure a subject from a privacy invasion attack using a physical stimulus. We show how a cyber-shirt can be used as a means of triggering a physical stimulus in order to detect whether a subject is being tracked by a drone.
We connected a LED strip to a microcontroller (Arduino Uno) and attached them both to a white shirt to create a cyber-shirt. We programmed the micorocontroller to modulate the pattern 01010011 01001111 01010011 (SOS in ASCII) using the LED strip as a light sequence (the physical stimulus). Again, we used on-off keying to modulate the binary sequence. Each bit was modulated using a window of five seconds. For 1 bits we flickered the LED strip, and for 0 bits we switched the LED strip off. A DJI Mavic was used by an operator  to record the person wearing the cyber-shirt. We repeated the same experimental setups used in the previous experiment. The video of the experiment was recorded by the DJI Mavic.

\begin{figure}
\centering
\includegraphics[width=1.00\columnwidth]{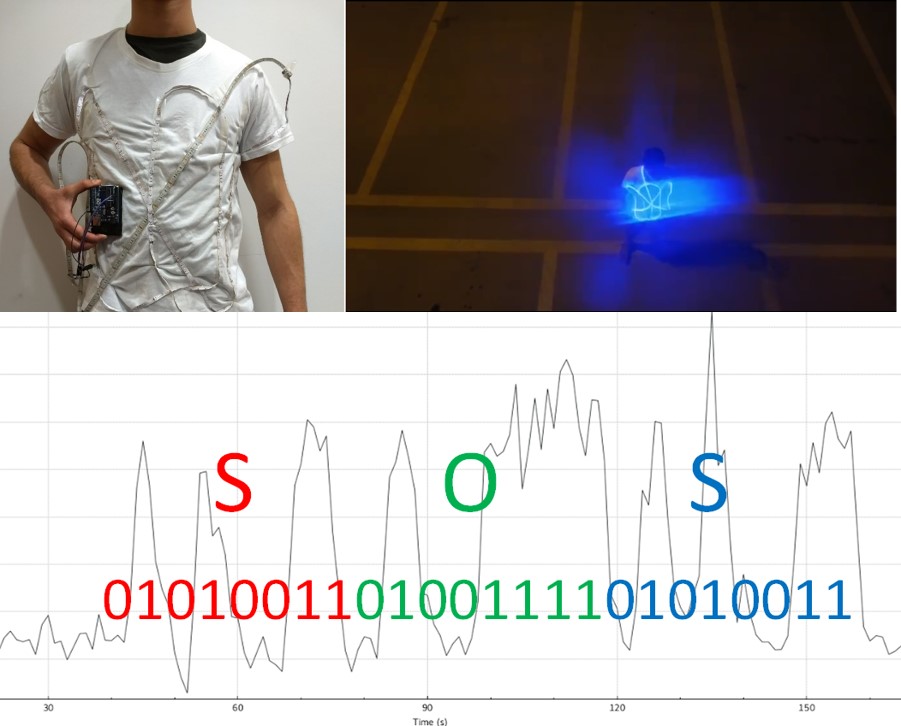}
\caption{Cyber-shirt experiment}
\label{fig:cyber-shirt}
\end{figure}

Figure \ref{fig:cyber-shirt} presents the results of applying Algorithm \ref{algorithm-lights} to the intercepted traffic. The peaks in the bitrate correlate to the time at which the LED strip was flickered. The flicker that was used to modulate the 1 bits increased the bitrate by 3-4 times, from an average of 20 KB to an average of 60-80 KB. As can be seen in Figure \ref{fig:cyber-shirt}c, the light sequence that was produced by the LED strip influenced the bitrate in a way that watermarked the bitrate to the given pattern that was programmed in the microcontroller.

\begin{figure}
\centering
\includegraphics[width=1.0\columnwidth]{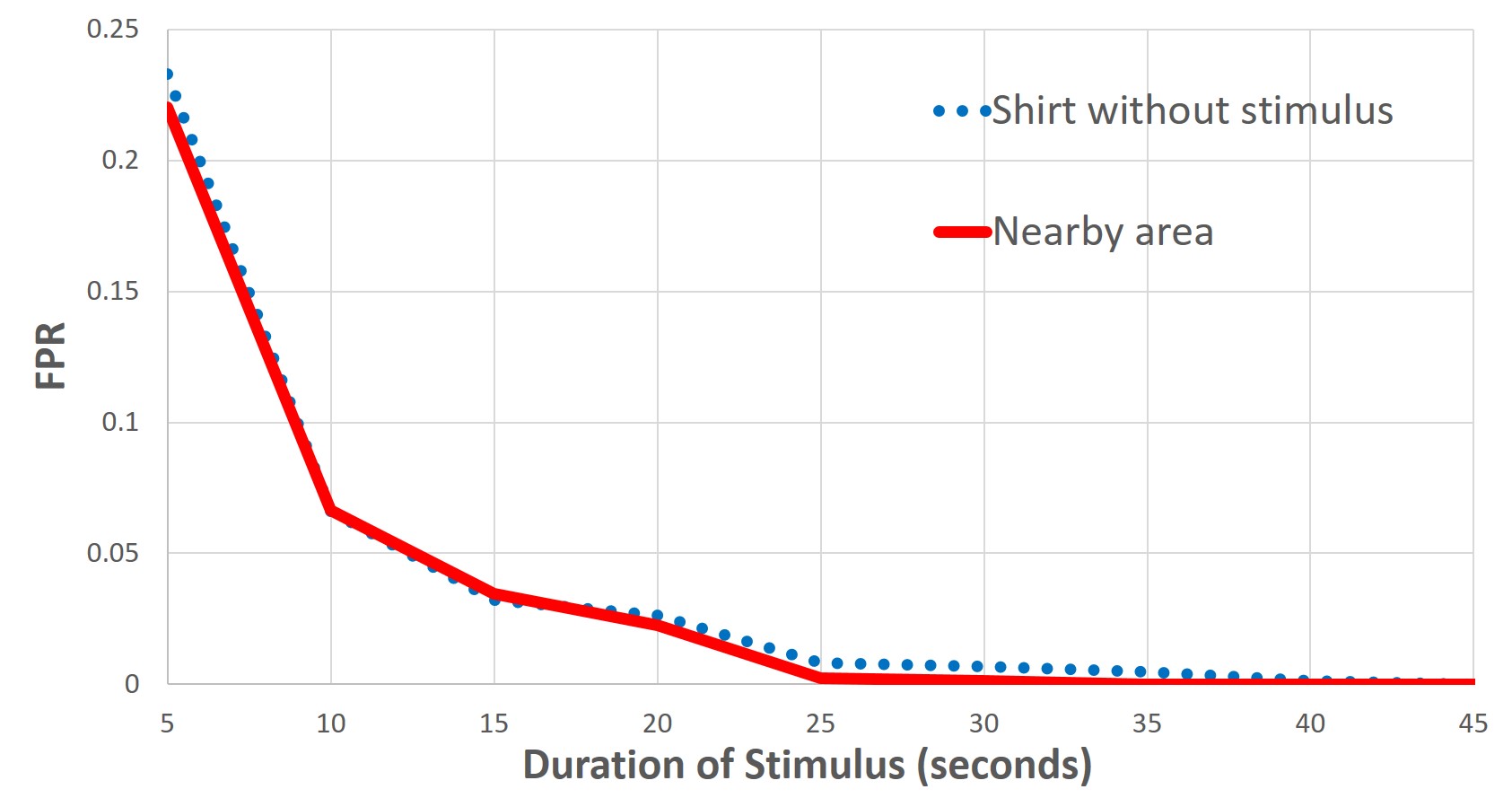}
\caption{A graph of the FPR as a function of the duration of the physical stimulus.}
\label{fig:ido-graph}
\end{figure}

In order to prove that the physical stimulus is the cause of the traffic change, we conducted one more experiment in which we streamed the same subject for 20 minutes without any physical stimulus conducted by the shirt. In addition, in order to prove that this effect could not be reproduced by another object nearby (that is not the target), we streamed the area for 20 minutes. In both cases we intercepted the traffic using the same experimental setup. We applied Algorithm \ref{algorithm-lights} to the intercepted traffic and calculated the false positive rate as a function of the duration of the physical stimulus for the original signal 01010011 01001111 01010011.  As can be seen from the results presented in Table \ref{tab:ido-fpr}, a pattern of 10 seconds is sufficient for excluding detection errors that occur when filming another target (with an FPR of 0.067). In addition,  a pattern of 10 seconds is sufficient for excluding errors that are coincidental and the result of wind or other physical movement (with an FPR of 0.066). The graph in Figure \ref{fig:ido-graph} presents the FPR as a function of the duration of the physical stimulus.

\begin{table}[]
	\centering
	\caption{FPR based on applying Algorithm \ref{algorithm-lights} on the same subject without a physical stimulus and another subject} 
    \resizebox{0.7\columnwidth}{!}{%
          \begin{tabular}{|l|l|l|}
    \toprule
    \multicolumn{1}{|p{4.19em}|}{\textbf{Duration of \newline{} Stimulus\newline{}(in seconds)}} & \multicolumn{1}{p{4.065em}|}{\textbf{The Subject\newline{}without Stimulus}} & \multicolumn{1}{p{3.565em}|}{\textbf{Nearby Area}} \\
    \midrule
    5     & 0.233 & 0.220 \\
    \midrule
    10    & 0.066 & 0.067 \\
    \midrule
    15    & 0.032 & 0.035 \\
    \midrule
    20    & 0.026 & 0.022 \\
    \midrule
    25    & 0.008 & 0.003 \\
    \midrule
    30    & 0.007 & 0.001 \\
    \midrule
    35    & 0.005 & 0 \\
    \midrule
    40    & 0.001 & 0 \\
    \midrule
    45    & 0     & 0 \\
    \bottomrule
    \end{tabular}%
	}
~\label{tab:ido-fpr}
\end{table}

\section{Countermeasures}

In this section we review countermeasures against privacy invasion attacks and more specifically, discuss countermeasures that can be used against our methods. 
In recent years several methods have been suggested to disable a drone. \textit{Son et al.} \cite{son2015rocking} presented a method to spoof the drone's gyroscope using ultrasound. \textit{Davidson et al.} \cite{davidson2016controlling} spoofed the drone's downward camera using a laser and projector that was projected on different surfaces. \textit{Luo et al.} \cite{Drones-Hijacking} showed different methods for hijacking and disabling a drone: (1) using GPS spoofing of no-fly zones, and during the return to home task, and (2) FPV jamming of the video stream. \textit{Rodday et al.} \cite{Hacking-A-Professional-Drone} showed techniques to hijack a \$30,000 drone by replaying maneuvering commands to the drone, and \textit{Kamkar et al.} \cite{SkyJack} performed a deauthentication attack on a Parrot AR.Drone.

In order to evade detection by an interceptor that uses our methods an attacker can use a drone equipped with two video cameras that are directed at different angles. The first camera will provide an FPV channel and be used by the operator for maneuvering. The second camera will be positioned 180 $^\circ$ from the first camera, and it will store the captured video to a memory card and be used by the operator to film a target. Using a drone with two cameras has a primary advantage and disadvantage. The primary advantage is that the second camera which is used for recording the target is immune to our detection methods, so an interceptor won't be able to detect the target POI by initiating a physical stimulus. However, using different cameras for maneuvering and for recording results in blind/semi-blind recording. Therefore the operator won't have total control of the specific object that is being recorded, because the FPV channel only streams the video from the first camera which is not directed at the target. Another method that can be used by an operator to evade detection is redundancy. In this case, the operator can program the drone to generate traffic at a fixed bitrate. This can be done by transmitting the raw data (instead of delta frames) that is captured by the camera at a low quality. Another option for implementing a fixed bitrate approach is to encode the captured data using delta-frames (in order to support high resolutions) and compensate for the difference in the bitrate between the B and P frames to the I-Frames with additional mock encrypted packets that will be sent from the drone in order and be discarded by the receiver. However, the main disadvantage of this method is that calculating the required volume of mock traffic to be sent will cause delays to the FPV channel.

\section{Conclusions}
\label{sec:Conclusions}

In this research we demonstrated methods that can be used to detect whether an object has been captured and is being streamed from a drone camera to its controller. While many methods have been suggested in recent years to detect the presence of a nearby drone, this research is the first to introduce methods that distinguish between the legitimate and illegitimate (for purposes of privacy invasion) use of a nearby drone. These days, consumer drones are used to conduct privacy invasion attacks throughout the world, however no tool currently exists for showing that a specific drone is being used to stream a target. Perhaps in the future, legislation will consider a watermarked PCAP as evidence, opening up new opportunities for implementing our methods in a system or service.

\section{Acknowledgments}
We would like to thank Roei Cohen and Ido Lavi for their help filming the videos.

\newpage

\bibliographystyle{IEEEtran}
\bibliography{IEEEabrv,main}

\begin{thebibliography}{10}
\providecommand{\url}[1]{#1}
\csname url@samestyle\endcsname
\providecommand{\newblock}{\relax}
\providecommand{\bibinfo}[2]{#2}
\providecommand{\BIBentrySTDinterwordspacing}{\spaceskip=0pt\relax}
\providecommand{\BIBentryALTinterwordstretchfactor}{4}
\providecommand{\BIBentryALTinterwordspacing}{\spaceskip=\fontdimen2\font plus
\BIBentryALTinterwordstretchfactor\fontdimen3\font minus
  \fontdimen4\font\relax}
\providecommand{\BIBforeignlanguage}[2]{{%
\expandafter\ifx\csname l@#1\endcsname\relax
\typeout{** WARNING: IEEEtran.bst: No hyphenation pattern has been}%
\typeout{** loaded for the language `#1'. Using the pattern for}%
\typeout{** the default language instead.}%
\else
\language=\csname l@#1\endcsname
\fi
#2}}
\providecommand{\BIBdecl}{\relax}
\BIBdecl

\bibitem{commercial-uav-market-analysis}
B.~Insider, ``Commercial unmanned aerial vehicle (uav) market analysis,''
  \url{http://www.businessinsider.com/commercial-uav-market-analysis-2017-8}.

\bibitem{Woman-grabs-gun-shoots-nosy-neighbour-s-drone}
D.~Mail, ``Woman grabs gun shoots nosy neighbour's drone,''
  \url{http://www.dailymail.co.uk/news/article-4283486/Woman-grabs-gun-shoots-nosy-neighbour-s-drone.html}.

\bibitem{Virginia-Woman-Shoots-Down-Drone-Near-Actor-Robert-Duvalls-Home}
N.~Washington, ``Virginia woman shoots down drone near actor robert duvalls
  home,''
  \url{http://www.nbcwashington.com/news/local/Virginia-Woman-Shoots-Down-Drone-Near-Actor-Robert-Duvalls-Home-391423411.html}.

\bibitem{not-my-backyard-man-arrested-after-shooting-drone-down}
N.~News, ``Kentucky man arrested after shooting down neighbor's drone,''
  \url{http://www.nbcnews.com/news/us-news/not-my-backyard-man-arrested-after-shooting-drone-down-n402271}.

\bibitem{cheating}
N.~Y. Post, ``Husband uses drone to catch cheating wife,''
  \url{https://nypost.com/2016/11/16/husband-uses-drone-to-catch-cheating-wife/},
  2016.

\bibitem{case-dismissed-against-william-h-merideth-kentucky-man-arrested-shooting}
N.~News, ``Case dismissed against william h. merideth, kentucky man arrested
  for shooting down drone,''
  \url{http://www.nbcnews.com/news/us-news/case-dismissed-against-william-h-merideth-kentucky-man-arrested-shooting-n452281}.

\bibitem{an-update-on-drone-privacy-concerns}
L.~360, ``An update on drone privacy concerns,''
  \url{https://www.law360.com/articles/848165/an-update-on-drone-privacy-concerns}.

\bibitem{you-cant-shoot-a-drone-so-what-can-you-do-if-it-invades-your-privacy}
N.~World, ``You cant shoot a drone so what can you do if it invades your
  privacy,''
  \url{http://www.networkworld.com/article/2941952/opensource-subnet/you-cant-shoot-a-drone-so-what-can-you-do-if-it-invades-your-privacy.html}.

\bibitem{Insurgents-Intercept-Drone-Video}
Wired, ``Insurgents intercept drone video in king-size security breach
  (updated, with video),''
  \url{https://www.wired.com/2009/12/insurgents-intercept-drone-video-in-king-sized-security-breach/}.

\bibitem{British-and-US-intelligence-hacked-into-Israeli-drones}
Telegraph, ``British and us intelligence 'hacked into israeli drones',''
  \url{http://www.telegraph.co.uk/news/worldnews/middleeast/israel/12128855/British-and-US-intelligence-hacked-into-Israeli-drones.html}.

\bibitem{russia-intercepted-us-drone-over-crimea}
B.~Insider, ``Russia says it intercepted a us drone over crimea,''
  \url{http://www.businessinsider.com/russia-intercepted-us-drone-over-crimea-2014-3}.

\bibitem{Why-Iran's-capture-of-US-drone-will-shake-CIA}
BBC, ``Why iran's capture of us drone will shake cia,''
  \url{http://www.bbc.com/news/world-us-canada-16095823}.

\bibitem{Nasrallah-Ynet}
Ynet, ``Nasrallah describes 1997 ambush,''
  \url{http://www.ynetnews.com/articles/0,7340,L-3932886,00.html}.

\bibitem{Nasrallah-Ynet2}
------, ``What really went wrong in botched 1997 shayetet 13 operation?''
  \url{http://www.ynetnews.com/articles/0,7340,L-4977429,00.html}.

\bibitem{saponas2007devices}
T.~S. Saponas, J.~Lester, C.~Hartung, S.~Agarwal, T.~Kohno \emph{et~al.},
  ``Devices that tell on you: Privacy trends in consumer ubiquitous
  computing.'' in \emph{USENIX Security Symposium}, 2007, pp. 55--70.

\bibitem{liu2010video}
Y.~Liu, A.-R. Sadeghi, D.~Ghosal, and B.~Mukherjee, ``Video streaming
  forensic-content identification with traffic snooping.'' in \emph{ISC}.\hskip
  1em plus 0.5em minus 0.4em\relax Springer, 2010, pp. 129--135.

\bibitem{203850}
\BIBentryALTinterwordspacing
R.~Schuster, V.~Shmatikov, and E.~Tromer, ``Beauty and the burst: Remote
  identification of encrypted video streams,'' in \emph{26th {USENIX} Security
  Symposium ({USENIX} Security 17)}.\hskip 1em plus 0.5em minus 0.4em\relax
  Vancouver, BC: {USENIX} Association, 2017, pp. 1357--1374. [Online].
  Available:
  \url{https://www.usenix.org/conference/usenixsecurity17/technical-sessions/presentation/schuster}
\BIBentrySTDinterwordspacing

\bibitem{WiFi-FPV-vs-5.8GHz-FPV-vs-2.4GHz-FPV-Ultimate-Guide}
rcdronearena, ``Wifi fpv vs 5.8ghz fpv vs 2.4ghz fpv: Ultimate guide,''
  \url{http://www.rcdronearena.com/2016/03/15/wifi-fpv-vs-5-8ghz-fpv-vs-2-4ghz-fpv-explained/}.

\bibitem{Wifi-FPV-Drones}
auselectronicsdirect, ``Wifi fpv drones,''
  \url{https://www.auselectronicsdirect.com.au/drones/fpv-drone/wifi-fpv-drones/}.

\bibitem{8-smartphone-controlled-drones}
androidauthority, ``8 fun drones you can control with your smartphone,''
  \url{https://www.androidauthority.com/best-smartphone-controlled-drones-744632/}.

\bibitem{8-DRONES-THAN-CAN-BE-CONTROLLED-BY-A-SMARTPHONE}
dronesglobe, ``8 drones than can be controlled by a smartphone (fully or
  partially),'' \url{http://www.dronesglobe.com/guide/smartphone-drones/}.

\bibitem{Spark-Remote-Controller}
DJI, ``Spark remote controller,''
  \url{https://store.dji.com/product/spark-remote-controller}.

\bibitem{PARROT-SKYCONTROLLER}
Parrot, ``Parrot skycontroller,''
  \url{https://www.parrot.com/global/accessories/drones/parrot-skycontroller#parrot-skycontroller}.

\bibitem{Phantom-range-extenders}
droneuplift, ``Top 5 best dji phantom signal range boosters 2017,''
  \url{http://www.droneuplift.com/top-5-dji-phantom-signal-range-extenders/}.

\bibitem{wiegand2003overview}
T.~Wiegand, G.~J. Sullivan, G.~Bjontegaard, and A.~Luthra, ``Overview of the h.
  264/avc video coding standard,'' \emph{IEEE Transactions on circuits and
  systems for video technology}, vol.~13, no.~7, pp. 560--576, 2003.

\bibitem{ostermann2004video}
J.~Ostermann, J.~Bormans, P.~List, D.~Marpe, M.~Narroschke, F.~Pereira,
  T.~Stockhammer, and T.~Wedi, ``Video coding with h. 264/avc: tools,
  performance, and complexity,'' \emph{IEEE Circuits and Systems magazine},
  vol.~4, no.~1, pp. 7--28, 2004.

\bibitem{jack2011video}
K.~Jack, \emph{Video demystified: a handbook for the digital engineer}.\hskip
  1em plus 0.5em minus 0.4em\relax Elsevier, 2011.

\bibitem{mitrovicvideo}
D.~Mitrovic, ``Video compression,'' \emph{University of Edinburgh}.

\bibitem{reed2016leaky}
A.~Reed and B.~Klimkowski, ``Leaky streams: Identifying variable bitrate dash
  videos streamed over encrypted 802.11 n connections,'' in \emph{Consumer
  Communications \& Networking Conference (CCNC), 2016 13th IEEE Annual}.\hskip
  1em plus 0.5em minus 0.4em\relax IEEE, 2016, pp. 1107--1112.

\bibitem{reed2017identifying}
A.~Reed and M.~Kranch, ``Identifying https-protected netflix videos in
  real-time,'' in \emph{Proceedings of the Seventh ACM on Conference on Data
  and Application Security and Privacy}.\hskip 1em plus 0.5em minus 0.4em\relax
  ACM, 2017, pp. 361--368.

\bibitem{wright2007language}
C.~V. Wright, L.~Ballard, F.~Monrose, and G.~M. Masson, ``Language
  identification of encrypted voip traffic: Alejandra y roberto or alice and
  bob?'' in \emph{USENIX Security Symposium}, vol.~3, 2007, pp. 43--54.

\bibitem{wright2008spot}
C.~V. Wright, L.~Ballard, S.~E. Coull, F.~Monrose, and G.~M. Masson, ``Spot me
  if you can: Uncovering spoken phrases in encrypted voip conversations,'' in
  \emph{Security and Privacy, 2008. SP 2008. IEEE Symposium on}.\hskip 1em plus
  0.5em minus 0.4em\relax IEEE, 2008, pp. 35--49.

\bibitem{white2011phonotactic}
A.~M. White, A.~R. Matthews, K.~Z. Snow, and F.~Monrose, ``Phonotactic
  reconstruction of encrypted voip conversations: Hookt on fon-iks,'' in
  \emph{Security and Privacy (SP), 2011 IEEE Symposium on}.\hskip 1em plus
  0.5em minus 0.4em\relax IEEE, 2011, pp. 3--18.

\bibitem{7417767}
C.~Wampler, S.~Uluagac, and R.~Beyah, ``Information leakage in encrypted ip
  video traffic,'' in \emph{2015 IEEE Global Communications Conference
  (GLOBECOM)}, Dec 2015, pp. 1--7.

\bibitem{eshel2013mobile}
T.~Eshel, ``Mobile radar optimized to detect uavs, precision guided weapons,''
  \emph{Defense Update}, 2013.

\bibitem{rozantsev2015flying}
A.~Rozantsev, V.~Lepetit, and P.~Fua, ``Flying objects detection from a single
  moving camera,'' in \emph{Proceedings of the IEEE Conference on Computer
  Vision and Pattern Recognition}, 2015, pp. 4128--4136.

\bibitem{busset2015detection}
J.~Busset, F.~Perrodin, P.~Wellig, B.~Ott, K.~Heutschi, T.~R{\"u}hl, and
  T.~Nussbaumer, ``Detection and tracking of drones using advanced acoustic
  cameras,'' in \emph{Unmanned/Unattended Sensors and Sensor Networks XI; and
  Advanced Free-Space Optical Communication Techniques and Applications}, vol.
  9647.\hskip 1em plus 0.5em minus 0.4em\relax International Society for Optics
  and Photonics, 2015, p. 96470F.

\bibitem{case2008low}
E.~E. Case, A.~M. Zelnio, and B.~D. Rigling, ``Low-cost acoustic array for
  small uav detection and tracking,'' in \emph{Aerospace and Electronics
  Conference, 2008. NAECON 2008. IEEE National}.\hskip 1em plus 0.5em minus
  0.4em\relax IEEE, 2008, pp. 110--113.

\bibitem{vasquez2008multisensor}
J.~R. Vasquez, K.~M. Tarplee, E.~E. Case, A.~M. Zelnio, and B.~D. Rigling,
  ``Multisensor 3d tracking for counter small unmanned air vehicles (csuav),''
  in \emph{Proc. SPIE}, vol. 6971, 2008, p. 697107.

\bibitem{peacock2013towards}
M.~Peacock and M.~N. Johnstone, ``Towards detection and control of civilian
  unmanned aerial vehicles,'' 2013.

\bibitem{birnbach2017wi}
S.~Birnbach, R.~Baker, and I.~Martinovic, ``Wi-fly?: Detecting privacy invasion
  attacks by consumer drones,'' \emph{NDSS}, 2017.

\bibitem{liu2008wavelet}
Y.~Liu, C.~Ou, Z.~Li, C.~Corbett, B.~Mukherjee, and D.~Ghosal, ``Wavelet-based
  traffic analysis for identifying video streams over broadband networks,'' in
  \emph{Global Telecommunications Conference, 2008. IEEE GLOBECOM 2008.
  IEEE}.\hskip 1em plus 0.5em minus 0.4em\relax IEEE, 2008, pp. 1--6.

\bibitem{Mavic-Pro}
DJI, ``Mavic pro,'' \url{https://www.dji.com/mavic}.

\bibitem{Intel-Dual-Band-Wireless-AC-8265}
Intel, ``Intel dual band wireless ac 8265,''
  \url{http://ark.intel.com/products/94150/Intel-Dual-Band-Wireless-AC-8265}.

\bibitem{Airmon-ng}
``Airmon-ng,'' \url{https://www.aircrack-ng.org/doku.php?id=airmon-ng}.

\bibitem{son2015rocking}
Y.~Son, H.~Shin, D.~Kim, Y.-S. Park, J.~Noh, K.~Choi, J.~Choi, Y.~Kim
  \emph{et~al.}, ``Rocking drones with intentional sound noise on gyroscopic
  sensors.'' in \emph{USENIX Security Symposium}, 2015, pp. 881--896.

\bibitem{davidson2016controlling}
D.~Davidson, H.~Wu, R.~Jellinek, V.~Singh, and T.~Ristenpart, ``Controlling
  uavs with sensor input spoofing attacks.'' in \emph{WOOT}, 2016.

\bibitem{Drones-Hijacking}
A.~Luo, ``Drones hijacking - multi-dimensional attack vectors and
  countermeasures,'' \emph{DefCon 24}.

\bibitem{Hacking-A-Professional-Drone}
N.~Rodday, ``Hacking a professional drone,'' \emph{Black Hat Asia}, 2016.

\bibitem{SkyJack}
S.~Kamkar, ``Skyjack,'' 2015.

\end{thebibliography}

\footnotesize 
\Urlmuskip=0mu plus 1mu\relax

\end{document}